\documentclass[graybox]{svmult}

\usepackage{mathptmx}        
\usepackage{helvet}          
\usepackage{courier}         
\usepackage{type1cm}         
\usepackage{makeidx}         
\usepackage{graphicx}        
\usepackage{multicol}        
\usepackage[bottom]{footmisc}
\usepackage{amsmath,amsfonts,amssymb}
\usepackage[all]{xy}
\usepackage{listings,color}
\usepackage{url}
\usepackage[noadjust]{cite}

\definecolor{backgroundColor}{rgb}{0.85,0.85,1}
\lstdefinelanguage{maple}{%
  morekeywords={ dummy, proc, begin, end, if, fi, then, else, do,%
                cmul, wedge, eval, subs, cmulGTensor, cmulTensor,%
                switch, gswitch, local, phi, evalm, asvd, cm, ID,%
                Id, CM, error, global, return, displayid, type,%
                Threads, Task, Continue, cmulW, cliterms,%
                Bsignature, elif, Start, clicollect, twist,%
                hinversegGrayCode, oplus, Walsh,%
                continuation,task,floor,add,cmulWpar,
                addUp,%
                dummy}
  morecomment=[l]{\#}
  }
\makeindex                   
\def\Clifford{\texttt{CLIFFORD}}
\def\Bigebra{\texttt{Bigebra}}
\def\Cliplus{\texttt{Cliplus}}
\def\Octonion{\texttt{Octonion}}
\def\Maple{\texttt{Maple}}
\def\Clical{\texttt{CLICAL}}
\def\CL{C\!\ell}

\def\code#1{\texttt{\small #1}}

\newcommand{\ed}{\end{document}}

\begin{document}
\title*{On Parallelizing the Clifford Algebra Product for \Clifford}
%
\titlerunning{On Parallelizing the Clifford Algebra Product for \Clifford}

%
\author{Rafa{\l} Ab{\l}amowicz and Bertfried Fauser}
\institute{Rafa{\l} Ab{\l}amowicz \at 
  Department of Mathematics, Box 5054,
  Tennessee Technological University,
  Cookeville, TN 38505, U.S.A. 
  \email{rablamowicz@math.tntech.edu}
\and
Bertfried Fauser \at 
  The University of Birmingham, 
  School of Computer Science,
  Edgbaston-Birmingham, B15 2TT, 
  England.
\email{b.fauser@cs.bham.ac.uk}}
%
%
\maketitle
\abstract*{--same abstract as below??--}
\abstract{We present, as a proof of concept, a way to parallelize the Clifford product in $\CL_{p,q}$ for a diagonalized quadratic form as a new procedure \code{cmulWpar} in the \Clifford\ package for
\Maple\textregistered. The procedure uses a new \code{Threads} module available under Maple 15 (and later) and a new \Clifford\ procedure \code{cmulW} which computes the Clifford product of any two Grassmann monomials in $\CL_{p,q}$ with a help of Walsh functions. We benchmark \code{cmulWpar} and compare it to two other procedures \code{cmulNUM} and \code{cmulRS} from \Clifford. We comment on how to improve \code{cmulWpar} by taking advantage of multi-core processors and multithreading available in modern processors.}
\section{Introduction} 
\label{sec:introduction} 
In~\cite{AF:submission1} we have described how to use (graded) tensor products and periodicity isomorphisms of real Clifford algebras to accomplish computations in Clifford algebras over vector spaces
of dimensions higher than~$8$. We accomplish these computations with Maple packages \Clifford\ and \Bigebra\ which have been described thoroughly 
in~\cite{A:1996, A:2009, AF:CLIFFORD,ablamowicz:fauser:2002d}. These packages have proven to be indispensable when deriving mathematical results presented in, for 
example,~\cite{AF:part1,AF:part2,AF:part3,AF:Hecke,ablamowicz:fauser:2002c}. Often \Clifford\ and \Bigebra\ have been used to prepare examples in support of a mathematical theory in place of hand computations, e.g.,~\cite{HHA:2012}, and especially when computing in higher dimensions.

Recent applications in engineering use real Clifford (geometric) algebras like $\CL_{8,2}$ when modeling geometric transformations in robotics.~\cite{bayro-corrochano:2012} Thus, there is a need for
efficient and fast symbolic computations which not only take advantage of the mathematical theory, for example by using the periodicity theorems, but take also full advantage of recent multi core hardware and software models supporting parallel computing.

In this note, we present an experimental procedure \code{cmulWpar} from \Clifford\ which utilizes the \emph{threading} module available in Maple 15 and later. Maple supports a coarse grained tasks based
model for parallel computing, which abstracts the need to actually deal with threads, locks and other low level constructs. The procedure \code{cmulWpar}, for now, computes, the Clifford product of two
arbitrary symbolic elements of type \code{clipolynom} in the real Clifford algebra $\CL_{p,q}$ for a diagonalized quadratic form.\footnote{In worksheets posted at~\cite{worksheets}, we show how to extend this multi-threading to quantum Clifford algebras $\CL(B)$ of any arbitrary bilinear form $B$.}

In \Clifford\ the user can chose between two algorithms to compute the Clifford product, or supply his/her own routine (not necessarily computing the Clifford product). The two main procedures are
\code{cmulNUM} and \code{cmulRS} which compute the Clifford product of any two basis monomials of type \code{clibasmon} in Clifford algebras $\CL(B)$ of any bilinear form $B$. The former is based on
Chevalley's recursive definition of the product and performs usually better for bilinear forms with numeric entries, especially if many of them are zero. The latter is based on the Hopf algebraic Rota-Stein cliffordization process and computes faster on fully symbolic bilinear forms. These routines are highly optimized for speed as they use internal features of Maple like hashing of already computed results (using the \code{remember} option). Although we have succeeded in parallelizing them after making all procedures internal to them thread-safe, in this note we concentrate on computations in real Clifford algebras $\CL_{p,q}$ of a non-degenerate quadratic form and the simpler \code{cmulWpar} procedure. 

A third experimental procedure available to \Clifford\ is \code{cmulW}. It belongs (for now) to \code{Walshpackage} developed by the authors. \code{cmulW} uses binary coding of basis elements and Walsh functions, see for example~\cite{lounesto:2001}, to compute the Clifford product of any two basis monomials in $\CL_{p,q}$ for a quadratic form of signature $(p,q)$ in an orthogonal basis. We like to recall that \Clical, a stand-alone semi-symbolic ``calculator" for $\CL_{p,q}$ designed by Lounesto \textit{et al.}~\cite{lounesto:1987,ablamowicz:sobczyk:2004} already in 1987, was based on binary coding and Walsh functions for internal data handling and storage. 

In Section~\ref{sec:code}, we display and briefly discuss the code of \code{cmulW} and \code{cmulWpar} which internally uses \code{cmulW} for a product of any two basis monomials. We describe a mechanism in \Clifford\ which permits the user to select which of the procedures \code{cmulNUM}, \code{cmulRS}, \code{cmulW}, or even a user provided procedure, is used internally by the active, non parallel, top-level procedure \code{cmul} furnishing the Clifford product in $\CL(B)$ (the first two) or $\CL_{p,q}$ (the third one). Then, we benchmark the procedures, namely, the parallel 
\code{cmulWpar} against the sequential \code{cmulW}, \code{cmul} with \code{cmulRS}, and \code{cmul} with \code{cmulNUM} for some test computations of the most general Clifford polynomials
in $\CL_{p,q}$ for $p+q\leq9.$ The complete and well commented code of all Maple worksheets showing these computations including parallelized \code{cmulNUM} and \code{cmulRS} is available
at~\cite{worksheets}.

\section{Code of \texttt{cmulW} and \texttt{cmulWpar}}
\label{sec:code}

\subsection{The Clifford product based on Walsh functions}
\label{sec:Walsh}
First, we present the code of \code{cmulW} which we use later in the parallel procedure \code{cmulWpar}. The latter procedure relies on several other procedures, which we do display here for the sake of
completeness, and which handle things like producing the Clifford product on basis monomials (\code{Walsh}) and the data conversion (\code{convert(<bas>,<data-type1>}) from \Clifford's internal data
structures for basis monomials and their representations as binary tuple used by the \code{oplus} and \code{Walsh} procedures. As \code{cmulRS} and \code{cmulNUM} do not have to perform these
conversions, there is a slight loss of speed here due to the data conversion. \code{twist} provides the proper sign factor due to the grading which is easily computed from the binary (Gray code)
representation of the Clifford monomials.

\begin{lstlisting}[caption={Clifford product on basis monomials \code{eI}, \code{eJ}
                            using Walsh functions in $\CL_{p,q}$},label={CST:Walsh}]
cmulW:=proc(eI::clibasmon,eJ::clibasmon,
            B1::{matrix,list(nonnegint)}) 
local a,b,ab,monab,Bsig,flag,i,dim_V_loc,ploc,qloc,
      _BSIGNATUREloc; 
# -- this procedure depends on external variables
global dim_V,_BSIGNATURE,p,q;
if type(B1,list) then
   ploc,qloc:=op(B1);
   dim_V_loc:=ploc+qloc:
   _BSIGNATUREloc:=[ploc,qloc]: 
   else 
   ploc,qloc:=p,q;   ###<<<-- this reads global p and q
   dim_V_loc:=dim_V: ###<<<-- this reads global dim_V
   _BSIGNATUREloc:=[ploc,qloc]:
   if not _BSIGNATURE=[ploc,qloc] then _BSIGNATURE:=[p,q] end if:
end if:
# -- data structure conversion: string to binary
a,b:=convert(eI,clibasmon_to_binarytuple,dim_V_loc),
     convert(eJ,clibasmon_to_binarytuple,dim_V_loc);
# -- mod 2 binary addition
ab:=oplus(a,b);
# -- data structure conversion: binary to string 
monab:=convert(ab,binarytuple_to_clibasmon);
return 
 twist(a,b,_BSIGNATUREloc)*Walsh(a,hinversegGrayCode(b))*monab;
end proc:
\end{lstlisting}

\subsection{Maple's threading mechanism for coarse grained parallel computing}
The following example is taken from Maple's help page \code{?Threads:-Task:-Start}.\footnote{See~\cite{maplesoft}.} It explains how to split a computation into pieces when the computation is `large' enough to profit from a parallel execution, and then execute the parallel tasks and use a continuation function to produce the result. The example computes $\sum_{i=1}^{10^7} i$.
\begin{lstlisting}[caption={Task threading example},label={lst:parexample}]
continuation := proc( a, b ) # add two results
   return a + b;
end proc;
task := proc( i, j )
   # distributes the computation into tasks
   local k;
   if ( j-i < 1000 ) then
       # if the range is small, just compute
       return add( k, k=i..j );
   else
       # split computation into two parts 
       k := floor( (j-i)/2 )+i;
       # produce two child tasks, by calling task recursively
       Threads:-Task:-Continue( continuation,
          Task=[ task, i, k ], Task=[ task, k+1, j ] );
   end if;
end proc;
# compute sum 1..10^7 parallel and using add
Threads:-Task:-Start(task,1,10^7) = add(i,i=1..10^7);
\end{lstlisting}
The parallelism is coarse-grained, the user does not have to deal with threads, and, for a large part, with locks. However, the involved routines have to be programmed in a thread-safe 
fashion.\footnote{The highly optimized routines \code{cmulRS} and \code{cmulNUM} turned up initially \textit{not} to be thread-safe due to relying on an internal function which was not defined in a thread-safe manner. After fixing this problem, both have been now successfully parallelized.} Since we want to demonstrate how to parallelize the Clifford product \code{cmulW} in $\CL_{p,q}$, in the next
section~\ref{sec:cmulWpar}, we will discuss only \code{cmulWpar}. The code of \code{cmulNUMpar} and \code{cmulRSpar} will be available in the worksheets~\cite{worksheets} accompanying this paper.

\subsection{The parallel procedure \texttt{cmulWpar} for the Clifford product}
\label{sec:cmulWpar}
We discuss briefly the code of \code{cmulWpar}, the parallelized version of the Clifford multiplication based on the Walsh functions core multiplication of the Clifford monomials used in \code{cmulW} shown in section~\ref{sec:Walsh}. This code needs at least Maple 15, and it has been tested in Maple 15 and Maple 16. 

The idea is to implement the Clifford multiplication along the lines of the example given above in the listing~\ref{lst:parexample}. As we parallelize a procedure with two arguments, we need to deal with each argument separately which slightly complicates the procedure.
\begin{lstlisting}[caption={Parallelized version \code{cmulWpar} of the Clifford 
product using \code{cmulW}},label={lst:cmulWpar}]
cmulWpar:=proc(x,y)
   local i,cf,term,co1,lst1,co2,lst2,lst,task,addUp;global p,q;
   # -- process x : turn Clifford polynomial x into a 
   #    list of type :: list [coeff,monom]
   if type(x,`+`) then    # x is a sum
     lst:=[op(x)];  lst1:=[]:
     for i in lst do      # split each term into [coeff,monom]
        if type(i,clibasmon) then
          lst1:=[op(lst1),[1,i]];
        else
          term,cf:=selectremove(type,i,clibasmon);
          lst1:=[op(lst1),[cf,term]];
        end if;
     end do;
   elif type(x,`*`) then  # x is a term
      term,cf:=selectremove(type,x,clibasmon);
      lst1:=[[cf,term]];
   else                   # x is a monom    
      lst1:=[[1,x]];
   end if;
   # -- process y :  turn Clifford polynomial y into a
   #    list of type :: list [coeff,monom]
   if type(y,`+`) then    # y is a sum
     lst:=[op(y)];
     lst2:=[]:
     for i in lst do      # split each term into [coeff,monom]
        if type(i,clibasmon) then
          lst2:=[op(lst2),[1,i]];
        else
          term,cf:=selectremove(type,i,clibasmon);
          lst2:=[op(lst2),[cf,term]];
        end if;
     end do;
   elif type(y,`*`) then  # y is a term
      term,cf:=selectremove(type,y,clibasmon);
      lst2:=[[cf,term]];
   else
      lst2:=[[1,y]];      # y is a monom
   end if;
   #====================================================
   # -- set up multitasking
   # -- continue function, add up results of task processes    
   addUp:=proc(a,b) a+b end proc:
   # -- task definition 
   task:=proc(lst1,lst2)
      local i,j, packsize,lst,lstpair;
      packsize:=4;        # size for sequential processing  
      # -- if x and y are small, just compute
      if max(nops(lst1),nops(lst2))<=packsize then
         add(add(lst1[i][1]*lst2[j][1]*cmulW(lst1[i][2],
             lst2[j][2],[p,q]),i=1..nops(lst1)),j=1..nops(lst2));
      # -- split the larger list for parallel processing
      elif nops(lst1) < nops(lst2) then   #    process lst2
         # -- split lst2 (y)
         lstpair:=lst2[1..packsize],lst2[packsize+1..-1];
         # -- produce two new tasks for the split list lst2
         Threads:-Task:-Continue(addUp, 
                 Task=[ task, lst1, lstpair[1]],
                 Task=[ task, lst1, lstpair[2]] );
      else   #    process lst1
         # -- split lst1 (x)
         lstpair:=lst1[1..packsize],lst1[packsize+1..-1];
         # -- produce two new tasks for the split list lst1
         Threads:-Task:-Continue(addUp,
                 Task=[ task, lstpair[1], lst2],
                 Task=[ task, lstpair[2], lst2] );
      end if;
   end proc:
   # -- start computation and collect results
   Threads:-Task:-Start( task, lst1,lst2);
end proc:
\end{lstlisting}
The procedure \code{cmulWpar} starts of by processing the inputs \code{x,y}, which may be Clifford polynomials. First, it splits \code{x} into a list \code{lst1} of lists of type 
\code{::List [coeff,monom]}, where \code{coeff} as a base ring element, and \code{monom} is a Clifford basis monomial $e_I$. This splitting is done using the type \code{clibasmon} from \Clifford. Similarly, \code{y} is split into the list \code{lst2}. This conversion could be made external by defining a new (external) procedure, however, as \Clifford\ deals internally differently with
(multi)linearity we keep it inline here, saving also two function calls. The signature $(p,q)$ of the current quadratic form is passed on to \code{cmulW} through two Maple variables $p$ and $q$ which are declared ``global" to \code{cmulWpar}.

The parallel processing starts with the definition of \code{addUp}, which adds the results later provided by two tasks. The main routine is \code{task}, which operates on pairs of type 
\code{:: List [coeff,monom]}. We pass here Cartesian products of the two lists in effect. Maple provides in the \code{combinat} package a way to pass an iterator, which saves memory. However, regarding thread safety we refrained from using this device yet. The parameter \code{packsize} sets a threshold from which size onwards parallel processing is applied. If both lists are small compared to
\code{packsize}, \code{task} just computes the result directly, as in the summation example in listing~\ref{lst:parexample}. Otherwise, one of the lists is `large' and we split the larger list recursively to produce two new tasks. To do so we use the \code{Threads:-Task:-Continue(...)} function. This proceeds unless both lists are `small' and are actually computed in their respective threads. Finally, the \code{Threads:-Task:-Start(...)} routine initializes the threading mechanism and starts producing the \code{task} in separate threads and also collects the results. 

The number of tasks produced is also the number of threads Maple produces. On a 4-core cpu one would like to have 4 threads only, all taking equal time to compute. For that reason we should compute the
parameter \code{packsize} dynamically. The Maple procedures like \code{Add}, \code{Seq}, etc., do this. At the moment we use a static \code{packsize} and have to compromise between an optimal number of
threads and losing parallelism at all. Experiments show that the input and dimension of $\CL$ have a large impact on a good choice for \code{packsize}, a reasonable setting is about 16. However, let
us consider two Clifford polynomials \code{x,y} with 1,000 terms each. Then the above procedure with \code{packsize=16} will produce roughly 3,906 tasks and hence as many threads. This clearly contradicts the idea of a coarse grained parallelism and calls for a dynamical setting of \code{packsize}.

As long as all involved procedures are thread-safe, that is, they can be used without any further (negative) side effects at a threat of miscomputing, parallelizing Maple procedures is formally straightforward. However, to achieve efficiency one needs some understanding of what is going on internally.

\section{Benchmarking results for \texttt{cmulWpar} versus sequential
multiplications \texttt{cmulW}, \texttt{cmulRS} and \texttt{cmulNum}}
\label{sec:benchmarks}
Benchmarking Maple procedures is not an easy task, and it becomes even more complicated if threads and parallel computing are in use. Firstly, Maple has a garbage collection and it is largely out of control when it does that. Using `cputime' as a measure, one needs to take into account that Maple adds up the cpu times on different cores. So, running a procedure on 2 cores for 3 seconds each will be reported 6 seconds of the cpu time usage. Having the administrative overhead, parallel computations will take \emph{more} cputime than single threaded computations. 

The second possibility is to use `realtime' which measures the clock time when executing a process. If the two processes above run in parallel, then this should take 3 seconds, but Maple has to share the processor with the operating system and possibly other applications which currently run. Hence, benchmarking has to be done on a clean idle system to get reproducible and comparable results, and this is what we have ensured to be the case. A useful tool for such benchmarking is the \code{CodeTools:-Usage(...)} procedure of Maple. However, the user may be warned that \code{profile} and 
\code{CodeTools} packages of Maple are \textit{not} yet thread safe and especially \code{profile} shows at times even negative run times.

We have tested the above given parallelized Clifford product on two machines. The first one is a dual core Windows XP (SP3) machine with Intel (R) Core (TM) 2 Duo CPU 2.19 GHz and 2.9 GB RAM. The second
machine is a core i7-2640QM at 2.8-3.5 GHz and 8 GB RAM running ubuntu 11.10 Linux and it has a physical dual core with 4 hyper threading virtual cores seen by linux. The two machines give, up to a scaling factor of about 2, the same results, so we show only one set of data here. See the appendix for Maple's perception of how many cores are available.

Having a dual or quad core available, we can expect a theoretical speedup of at most a factor of 2 or 4, which in practice suffers from administrative overhead of the threading software. The below given speedups cannot then be attributed to the parallelizing alone, but must also be partially caused by the different ways the procedures compute and by the involved data structures. It looks like as if thinking about the threading model leads one to more efficient code per se. A sort of a measure to check if parallelization does really occur is given by 
\begin{itemize}
\item[a)] checking the system load during computation to see if all cores are running under full load, 
\item[b)] computing the quotient cputime/realtime, which somehow measures the `effective number of cores in use'. 
\end{itemize}

The ratio cputime/realtime varies a lot over the input to \code{cmulWpar} and it seems to vary from 0.9 to 1.8 on the second machine given above. Maple allows one to set the number of cores (called cpu's in Maple) in use. In this way one can, with care (see the appendix) benchmark the parallelized threaded code on a single core versus several cores. As we do not have quad and octo-core machines available, we cannot demonstrate such results which would really show how the threading mechanism scales on the number of cores.

In Table~\ref{tab:bench} we summarize some of our benchmarking results.\footnote{These times were obtained on Windows XP (SP3) machine with Intel (R) Core (TM) 2 Duo CPU 2.19 GHz and 2.9 GB RAM.}
We have computed Clifford products of two most general Clifford polynomials $X$ and $Y$ in the Clifford algebras $\CL(p,0)$ for $p\leq 9$. The table shows CPU times in seconds as they were returned by the Maple procedure \code{CodeTools:-Usage(...)} where $t_1$ is the CPU time taken to compute such product by the parallel procedure \code{cmulWpar}, whereas $t_2$, $t_3$, and $t_4$ are the cpu times needed by the ordinary, non-parallel Clifford product procedure \code{cmul} using internally \code{cmulW}, \code{cmulRS}, and \code{cmulNUM}, respectively. The computations in dimensions 8 and 9 with \code{cmulW}, \code{cmulRS}, or \code{cmulNUM} were not completed due to running out the RAM memory. For example, in dimension 9 they were stopped after 7.8GB RAM had been consumed and the machine started to swap.


\begin{table}[t]
\caption{Benchmarking CPU times: $t_1$ of \code{cmulWpar};\newline 
$t_2$ of \code{cmulW}, $t_3$ of \code{cmulRS} and $t_4$ of \code{cmulNUM}}
\label{tab:bench}
\renewcommand{\arraystretch}{1.3}
\begin{tabular}{|c|c|c|c|c|c|c|c|}\hline
$\dim V$ & $t_1 [sec]$ & $t_2 [sec]$ & $t_2/t_1$ & $t_3 [sec]$ & $t_3/t_1$ & $t_4 [sec]$ & $t_4/t_1$\\\hline
2 & $0$       & $0.063$ & $\infty$ & $0.047$ & $\infty$ & $0.047$ & $\infty$\\\hline
3 & $0.046$   & $0.172$ & $3.7$    & $0.250$ & $5.44$   & $0.203$ & $4.41$\\\hline
4 & $0.250$   & $0.766$ & $3.1$    & $0.828$ & $3.31$   & $0.984$ & $3.93$\\\hline
5 & $1.0.94$  & $3.000$ & $2.74$   & $4.594$ & $4.20$   & $5.532$ & $5.05$\\\hline
6 & $4.687$   & $14.68$ & $3.13$   & $36.09$ & $7.7$    & $43.61$ & $9.3$\\\hline
7 & $22.000$  & $136.0$ & $6.18$   & $506.1$ & $22.9$   & $448.6$ & $22.5$\\\hline
8 & $112.47$ & \mbox{NA} & \mbox{NA} & \mbox{NA} & \mbox{NA} & \mbox{NA} & \mbox{NA}\\\hline
9 & $647.20$ & \mbox{NA} & \mbox{NA} & \mbox{NA} & \mbox{NA} & \mbox{NA} & \mbox{NA}\\\hline
\end{tabular}
\end{table}

\section{Conclusions}
\label{sec:conclusions}
We have shown how to use Maple's task model, a coarse-grained parallel computing framework, to parallelize the Clifford product for the \Clifford\ package. As long as such computations are side-effects free and thread safe, this is easily achieved by using the \code{Threads} package from Maple.

We hit on some problems when trying to parallelize the core multiplication procedures \code{cmulRS} and \code{cmulNUM}, which are highly optimized mathematically in their algorithms, and, on the software side, by extensive hashing of precomputed results (Maple remember tables). However, after removing some internal procedures these are now thread safe too, and have now parallel versions 
\code{cmulNUMpar} and \code{cmulRSpar}.

In this paper we have discussed the very fast procedure \code{cmulW} for computing the Clifford product in orthogonal bases and arbitrary signature as it was already thread-safe and allowed immediate parallelization. Amazingly the parallelized version is \textit{much faster} than the theoretical limit allows, so that this speedup is not solely due to our parallelizing the computation. It seems that dealing with the threading package of Maple forced us to produce more efficient coding especially of the multilinear features. We have provided detailed benchmark results showing the speedups and also discussed how to separate the speedup by the given different coding and that coming from actual parallel computing. We find a speedup on large Clifford polynomials due to parallelizing of up to
1.8 on a dual core machine, which is what one can expect and shows that parallel computing is by now feasible in symbolic computer algebra. This is possibly good news for engineers and roboticists who do computations in higher dimensional Clifford algebras like $\CL_{8,2}$ for geometric computations. 

Thus, we must be cautious when examining the speedup ratios $t_4/t_1$ and $t_3/t_1$ shown in Table~1. We repeat to caution the reader that, for example, the speedup factor of around 22 in dimension 7 cannot be attributed exclusively to the parallelization process alone. This speedup is a combination of factors, due to, for example, making the overhead in \code{cmulWPar} dealing with the bilinearity much smaller and faster than the resources- and time-demanding procedure \code{clibilinear} from \Clifford. The latter includes among other things time-consuming type checking of the input on many levels of recursion, especially when \code{cmulNUM} is used. At the moment we do actually profit from multi-threading seen by computing the number of effective cores 
$\mbox{cputime/realtime} \simeq 0.9\mbox{--}1.5$. However, we saw that a bit of reorganization of the data structures and the recursive way to do the products (saving memory) gives us an even
more substantial speedup. It is questionable if multi-threading is only valuable when one knows that one's code is at its theoretical limit with respect to space and time complexity, and \Clifford\ is not yet at that limit. We suspect that \Clifford\ could be faster at least by an overall factor of more than 20-30, based on this current experience, by a generic rewrite using better data structures and avoiding all the repetitious parsing and type checking where it can be avoided, and using the recursive way to split (multi)linearity, etc. Optimizing \Clifford\ and its related packages like \Bigebra, \Cliplus, \Octonion, etc. \cite{AF:CLIFFORD} is a priority whose urgency has been emphasized by this exercise in parallelizing the Clifford product.

The results discussed here are accompanied by Maple worksheets posted on~\cite{worksheets}. These well-documented worksheets contain further results and alternatives, as using the inherently parallel
procedures \code{Add}, \code{Seq}, \code{Map} of Maple or directly producing threads. There we further discuss the efficient usage of Maple's \code{Threads} package. We are working to make all of
\Clifford\ thread safe after we have succeeded parallelizing the more complex and complicated \code{cmulRS} and \code{cmulNUM} routines. While \code{cmulRS} is based on a provable optimal algorithm, the
above discussion still sheds some light on efficiency of the implementations due to different data structures or recursive computing models (saving memory usage). In that respect, this is a very open area of research.

\begin{acknowledgement}
Bertfried Fauser wants to thank Darin Ohashi from Maplesoft for his kind help with and email discussions about Maple's threading mechanism. Both authors thank referees for their comments as they have helped us extend this work and improve its clarity.
\end{acknowledgement}
\section*{Appendix}
\addcontentsline{toc}{section}{Appendix}
In this appendix we make some remarks about programming practice using the different versions of parallelizing code in Maple, using the threading module of Maple (from version 15 onwards), and the inherently parallel routines such as \code{Add}, \code{Mul}, \code{Map}, and \code{Seq}.

\subsection{Thread safety}
The \code{Threads} package was introduced in Maple 15 and it was improved for Maple 16. Still, large parts of Maple are not yet `thread safe', that is, the code cannot be run in parallel as it may cause side effects which can interfere with other threads. A common source of such problems are global variables and name space conflicts. For example, Maple's parser will not complain when a running variable in \code{add(f(i),i=1..10);} or \code{seq(i\^{}2,i=1..10);} is not declared local: it will simply miscompute. The following procedure \code{dummy} will miscompute in a threading
environment unless the local variables \code{i} and \code{j} are explicitly declared local to \code{dummy}.
\begin{lstlisting}[caption={Local running variables \textbf{have} to be declared local}]
dummy:=proc(x::List[expression],N::Integer)
   # local i,j;   # <== needed!
   # assignment of j produces a waring if not declared local
   j:=x[1];
   add(x[i],i=1..N);
end proc:   
\end{lstlisting}
Note, that Maple's parser does \textbf{not} complain about an undeclared local variable, so this issue slips through unnoticed, as it will for the (here unused) variable \code{j} if not declared local.

The next issue is more subtle. If in a procedure \code{dummy} one has a helper procedure \code{fun}, declared local to it, and if this function is defined with a remember table, either by option or by assigning certain values to it, then these assignments are seen globally, hence are visible to all threads! In effect, any thread can access values set other threads, and this will ultimately lead to errors.
\begin{lstlisting}[caption={\textbf{Avoid} procedures local to a procedure}]
# NOT thread safe procedure
dummy:=proc(x::List[expression],N::Integer)
   local fun,i;
   # -- local helper procedure
   fun:=proc(y::any) 
      # -- either remember here
      option remember;
      return 3*y^2+4; # just some computation
   end proc:
   # -- or use special cases (implements a remember table)
   fun(x[1]):=x[1];
   fun(x[N]):=x[N];
   add(fun(x[i]),i=1..N);
end proc:   
\end{lstlisting}
\Clifford\ used such a construction to implement permutations in reordering wedge products of Grassmann basis monomials, and this rendered \code{cmulNUM} and \code{cmulRS} not thread safe at first. The
reordering had to be done sequentially or it needed to be done differently.

A further issue with threading comes from the fact that the programmer must check \emph{every} procedure the given package uses whether it is `thread safe'. Each Maple procedure has a help page and there
it is marked if this particular procedure is thread safe, and for which version of Maple onwards. If no such statement is given, one must assume that the procedure as not thread safe. A suggestion of the referee to use the \code{combinat:-cartprod} construction to iterate over Cartesian products of sets cannot be realized, as this function is \textit{not} (yet) declared by Maple thread safe.

Unfortunately for benchmarking issues, neither \code{profile} nor the \code{CodeTools} package is thread safe yet either. While \code{profile} seems to be broken as it reports once in a while negative running times, the function \code{CodeTools:-Usage} seems to be reasonably stable. For that reason we used this utility to do our benchmarks, but ran also checks on the results via external timing.

\subsection{Overhead versus gain in parallel code}
Given the example from listing~\ref{lst:parexample} one observes that just adding integers in parallel is \emph{slower} than doing so sequentially. A similar result is obtained if one uses the parallel code snippet \code{Add(i,i=1..10\^{}7);}. This shows clearly that the overhead introduced by Maple to produce threads is \textit{too large} to provide any gain in speed. However, the situation changes when one computes a more complicated sum, e.g., $\sum_{i=1}^{10^7} i^{2/3}$ evaluated as float. The code snippet looks like \code{Add(evalf(i\^{}(2/3)),i=1..10\^{}7);} Similar effects are encountered when one uses threads directly. To benefit from parallelizing code in Maple, one has to make sure that the work done in a single task is \textit{as large as possible}, and that reflects the idea of
coarse grained parallelism implemented by Maple. 

Another suggestion of the referee to use \code{Add}, \code{Mul}, \code{Seq}, or \code{Map} \emph{inside} a thread is also not advisable. On a processor with $n$ cores (cpus), once $n$ threads already have been produced by other devices, parallelizing will only create superficial threads which cannot be processed in parallel since all cpus are already busy running assigned to them threads.

Finally, there is a difference in the \code{Threads} packages for Maple 15 and 16 how the number of cpus is set. The procedure \code{kernelopts(numcpus);} reports the number of `cpus' Maple sees and
uses for threading. Maple 15 sets the number of cpus to the number of virtual cores of a physical cpu (for example, 4 for a core i7-2640QM with 2 physical and 4 virtual hyper-threading cores), while Maple 16 uses the number of physical cores, that is 2 here. However, on modern core i7 processors the virtual cores allow a substantial speed up due to interlocking processes when queues and/or pipelines are filled etc. It is therefore advisable to set the number of cpus at the \emph{beginning} of the worksheet to the number of virtual cores by \code{kernelopts(numcpus=4);} for 4 virtual cores. Note that when the threading mechanism is used, it is initialized using this number, and the number of cpus \emph{cannot} be reset later again (despite what \code{kernelopts(numcpus);} later reports). 

\subsection{Other parallelization mechanisms of Maple}
As we have already mentioned, the creation of \emph{tasks} is not the only way you can use Maple's threading package. We have investigated if different approaches give significantly different results.

A first group of seemingly simple to use procedures are \code{Add}, \code{Mul}, \code{Seq}, and \code{Map}, which parallelize the corresponding sequential (lower case) procedures. The advantage is 
that Maple does some load balancing in computing how many threads are created and that it is very simple to use these procedures. As the previous section shows, one is nevertheless left with 
benchmarking these procedures as a too naive usage may result in slower code. Given the Clifford product we were able to get roughly a similar speedup as with the method described above, which does not 
yet use a dynamical setting of \code{packsize}. But using \code{Add} to sum up the terms in a Clifford product is very memory intensive, and this favors other solutions.

The second method is to use the task method, using the \code{Threads:-Task} package providing the procedures \code{Start}, \code{Return} to leave a task, and \code{Continue}. As we have chosen this model above, there is not much to add here.

A third way to create threads is to directly use the \code{Threads:-Create} command. Maple provides locks, mutexes and a synchronization using \code{Threads:-Wait} to deal with these threads directly. For example, to compute the sum of integers $\sum_{i=1}^{10^7}$ in 2 threads one can use this code:
\begin{lstlisting}
restart:
# -- define two _functions_ performing the work
p1:=proc(x) local i; add(i,i=1       ..5*10^6) end proc:
p2:=proc(x) local i; add(i,i=5*10^6+1..10^7  ) end proc:
# -- create 2 threads executing the work
id1:=Threads:-Create(p1(),out1);
id2:=Threads:-Create(p2(),out2);
# -- wait for the two threads id1, id2 to be finished
Threads:-Wait(id1,id2);
# produce the result
out1+out2;
\end{lstlisting}
The above code produces an output \code{id1=1}, \code{id2=2} --showing the id's of these threads--, and the numerical sum \code{50000005000000}. It is clear that this gives the most direct access to the threading mechanism, as the programmer can decide explicitly anything about the threads. This model is especially useful when several threads have to share resources, as one can lock variables, etc., using the \code{ConditionVariable} and \code{Mutex} packages. We have benchmarked a version of \code{cmulWpar} using this direct method (using numcpus=4 threads) and again obtained essentially the same performance results. Finding this, it seems to be advisable to use in Maple the easiest threading model available for the task at hand, as we have demonstrated.
\providecommand{\bysame}{\leavevmode\hbox to3em{\hrulefill}\thinspace}

\end{document}